# Fast and Robust Parametric Estimation for Time Series and Spatial Models

Stéphane Guerrier† and Roberto Molinari‡

**Abstract:** We present a new framework for robust estimation and inference on second-order stationary time series and random fields. This framework is based on the Generalized Method of Wavelet Moments which uses the wavelet variance to achieve parameter estimation for complex models. Using an M-estimator of the wavelet variance, this method can be made robust therefore allowing to estimate the parameters of a wide range of time series and spatial models when the data suffers from outliers or different forms of contamination. The paper presents a series of simulation studies as well as a range of applications where this new approach can be considered as a computationally efficient, numerically stable and robust method which performs at least as well as existing methods in bounding the influence of outliers on the estimation procedure.

**Keywords and phrases:** Generalized Method of Wavelet Moments, Spatial Models, Computational Efficiency.

## 1. Introduction

Parametric inference on random fields has been a widely tackled topic, especially concerning time series and spatial modelling. However, many available methods for this purpose can require a selection of auxiliary parameters (or models) which is not always clear, can be computationally impractical when dealing with larger sample sizes or can be numerically unstable when estimating complex models. Moreover, adding robust inference to this setting is often a daunting task. Indeed, already in the time series field (i.e. one-dimensional), the limited applicability of robust estimation is testified by the lack of available methods in statistical software, even though there is an abundance of literature in this domain. This is also evident when it comes to the robust estimation of spatial models. The reader can find a short literature review of existing methods for the robust estimation of random fields in Appendix A.

In this setting, the goal of this paper is to describe a robust method for the parametric estimation of time series and spatial models which is able to estimate a wide variety of second-order intrinsically stationary processes (i.e. stationary or non-stationary processes with stationary backward differences), is computationally efficient, numerically stable and compares on level, if not better in some cases, with the few already available robust methods. To do so, we first extend the idea of the Generalized Method of Wavelet Moments (GMWM) (see Guerrier et al., 2013) to the parametric estimation of random fields, having been initially conceived for time series model inference. The reason for this extension resides in the fact that the GMWM is based on a quantity called the Wavelet Variance (WV) which adequately summarizes the dependence structure "information" in a sample issued from a certain parametric model $F_{\boldsymbol{\theta}}$, with $\boldsymbol{\theta}$ being the parameter vector of interest. As a minimum distance estimator (MD), the GMWM uses the WV as auxiliary parameters that allow to estimate $\boldsymbol{\theta}$ for a variety of models which can also be relatively complex in nature (e.g. *latent* models which are the result of the sum of different underlying processes). Due to these characteristics, this method presents a few advantages over traditional estimators such as, for example, the Generalized Method of Moments (GMM) estimators and Maximum Likelihood Estimators (MLE). Indeed, for GMM estimators the selection of good moment conditions or auxiliary parameters becomes an important issue to deal with, especially when the sample size is large (see, for example, Andrews, 1999). On the other hand, when dealing with state-space time series models or simple spatial models, the MLE often relies on the state estimation, inversion of the covariance matrix and/or computation of the distance matrix thereby making their computational feasibility limited even when the sample is moderately large. With these issues in mind, the GMWM overcomes these problematics, paying a reasonable price in terms of statistical efficiency, by using a wavelet decomposition which adequately condenses the information in the sample to a moderate number of auxiliary parameters which are the WV and can consequently estimate also complex models in a computationally efficient manner.

Aside from the above advantages, the reason for considering the GMWM for the robust parametric estimation of time series and spatial models resides in the fact that it can easily be made robust by using a robust estimator of the WV. In fact, Ronchetti and Trojani (2001) and Genton and Ronchetti (2003) highlight that a bounded auxiliary parameter or moment condition can guarantee the robustness of the resulting parametric estimator and this property was already investigated in the time series setting in Guerrier et al. (2014) where simulation studies hinted that this approach constituted a valid means to bound the influence of contaminated observations in a dependent data scenario. However, in the latter the authors used the robust M-estimator of WV proposed by Mondal and Percival (2012b) that, although bounding the influence of outliers, does not benefit from clear asymptotic properties which would allow for inference when estimating parameters of random fields. For this reason, in this paper we make use of the M-estimator of WV described in Guerrier and Molinari (2016b) which does not necessarily require normality of the data, benefits from appropriate asymptotic properties and overall shows better finite sample performance. More specifically, this estimator also allows to determine the desired level of robustness and its consistency and joint asymptotic normality have been proven thereby ensuring that its properties directly transfer to the GMWM estimator.

This paper is organized in the following manner. Section 2 introduces the straightforward extension of the GMWM to the (robust) estimation of random fields and highlights its asymptotic properties. A simulation study comparing standard and robust estimators for the parameters of a variety of time series and spatial models is presented in Section 3 where the good properties of the Robust GMWM (RGMWM) are shown both in terms of estimation as in terms of computational efficiency and numerical stability. The usefulness of this new method is highlighted in Section 4 with a series of practical examples covering some applications in time series analysis as well as in spatial modelling. Finally, Section 5 concludes.

## 2. Robust GMWM for Random Fields

The GMWM was initially proposed by Guerrier et al. (2013) as a means to estimate so-called *composite* (or *latent*) time series models where the observed process is the result of a sum of different underlying processes which are intrinsically stationary. This allows to estimate, for example, a class of basic time series models as well as many linear state-space models among which we can find the classic (Seasonal) Autoregressive (Integrated) Moving Average (SARIMA) models (see for example the results of Granger and Morris, 1976). To do so, the GMWM uses the WV as an auxiliary parameter and is defined as

$$\hat{\boldsymbol{\theta}} = \operatorname*{argmin}_{\boldsymbol{\theta} \in \boldsymbol{\Theta}} (\hat{\boldsymbol{\nu}} - \boldsymbol{\nu}(\boldsymbol{\theta}))^T \boldsymbol{\Omega} (\hat{\boldsymbol{\nu}} - \boldsymbol{\nu}(\boldsymbol{\theta})) \qquad (2.1)$$

where $\boldsymbol{\theta}$ represents the vector of parameters defining the stochastic process $F_{\boldsymbol{\theta}}$, $\hat{\boldsymbol{\nu}} = [\hat{\nu}_j^2]_{j=1,\ldots,J}$ represents the vector of estimated WV (see Percival, 1995), $\boldsymbol{\nu}(\boldsymbol{\theta}) = [\nu_j^2(\boldsymbol{\theta})]_{j=1,\ldots,J}$ represents the vector of WV implied by the model of interest and $\boldsymbol{\Omega}$ is a weighting matrix, with elements $\omega_{i,j}$, $i,j = 1,\ldots,J$, chosen in a suitable manner (see Guerrier et al., 2013).

Having defined the GMWM, this section combines the M-estimation framework for the WV proposed in Guerrier and Molinari (2016b) to obtain the RGMWM estimator for multidimensional second-order intrinsically stationary random fields. From now onwards, the estimator of WV $\hat{\nu}_j^2$ therefore denotes the mentioned M-estimator and, based on this, in the following paragraphs we list and discuss the conditions for the consistency and asymptotic normality of the RGMWM which are generally the same as those for GMM estimators. Denoting $\boldsymbol{\nu}(\boldsymbol{\theta})$ as $\boldsymbol{\nu}$ for simplicity and using $N$ to represent the sample size, let us therefore start by listing the conditions for the consistency of the RGMWM estimator:

**(C1)** The set $\boldsymbol{\Theta}$ is compact.
**(C2)** The function $\boldsymbol{\nu}(\boldsymbol{\theta})$ is continuous in $\boldsymbol{\theta}$.
**(C3)** The function $\boldsymbol{\nu}(\boldsymbol{\theta})$ is globally identifiable.
**(C4)** $\|\hat{\boldsymbol{\nu}} - \boldsymbol{\nu}\| \xrightarrow{\mathcal{P}} 0$.
**(C5)** $\boldsymbol{\Omega}$ is positive definite.
**(C6)** If $\boldsymbol{\Omega}$ is estimated by $\hat{\boldsymbol{\Omega}}$, then $\hat{\omega}_{i,j} = \omega_{i,j} + \mathcal{O}_p(1/g(N))$ where $g(N) \in \mathbb{R}$ is a function of $N$ such that $J^2/g(N) = o(1)$.

Condition **(C1)** is a standard condition which is often used for estimators to be consistent while Condition **(C2)** is easy to verify and is respected for most intrinsically stationary processes. However, Condition **(C3)** is an essential one which is often hard to verify. In this case, the identifiability of a wide class of (latent) time series models and some spatial models was shown in Guerrier and Molinari (2016a) thereby verifying Condition **(C3)** for these cases. As for Condition **(C4)**, this is verified under the conditions presented for the M-estimator of WV given in Guerrier and Molinari (2016b). Generally speaking, according to whether a bounded or unbounded score function is chosen to obtain $\hat{\boldsymbol{\nu}}$, these conditions require the WV $\boldsymbol{\nu}$ to be identifiable and assume that the wavelet coefficients follow a strongly mixing stationary and ergodic process with bounded fourth moments. Moreover, if using a bounded score function, this must be Bouligand differentiable and its spectral density must be strictly positive at zero frequency. In addition, it is required that the number of scales of decomposition $J$ grow at a suitable rate compared to the sample size $N$ (see Guerrier and Molinari, 2016b). For multidimensional random fields (e.g. spatial processes), the last condition would require that the sample size goes to infinity along all dimensions at a suitable rate as well. Finally, Conditions **(C5)** and **(C6)** concern the choice of the weighting matrix $\boldsymbol{\Omega}$, where the first is easy to verify while the second does not appear strong to assume. Indeed, the GMWM estimator is consistent for any $\boldsymbol{\Omega}$ that is positive definite. However, the most efficient GMWM estimator is the one based on $\boldsymbol{\Omega} = \mathbf{V}^{-1}$ where $\mathbf{V}$ is the covariance matrix of the estimator of WV. Defining $\Psi(\mathbf{W}_{k_J}, \boldsymbol{\nu})$ as the vector score-function of the M-estimator proposed in Guerrier and Molinari (2016b), where $\mathbf{W}_{k_J}$ is the vector of wavelet coefficients at coordinates $k_J$, this covariance matrix is given by

$$\mathbf{V} = \mathbf{M}^{-1}\mathbf{S}_\psi(\mathbf{0})\mathbf{M}^{-T},$$

where $\mathbf{S}_\psi(\mathbf{0})$ is the spectral density of $\Psi(\mathbf{W}_{k_J}, \boldsymbol{\nu})$ at zero frequency and

$$\mathbf{M} = \mathbb{E}\left[-\frac{\partial}{\partial \boldsymbol{\nu}}\Psi(\mathbf{W}_{k_J}, \boldsymbol{\nu})\right].$$

We propose to estimate $\mathbf{V}$ either via parametric bootstrap or by replacing $\boldsymbol{\nu}$ in the above expressions with $\hat{\boldsymbol{\nu}}$ and estimating $\mathbf{M}$ and $\mathbf{S}_\psi(\mathbf{0})$ via their sample versions (see Iverson and Randles, 1989). With these conditions and denoting $\boldsymbol{\theta}_0$ as the true parameter vector of the random field $(X_k)$, we can now state the consistency of the GMWM estimator $\hat{\boldsymbol{\theta}}$.

**Proposition 2.1.** *Under conditions (C1) to (C6) we have that*

$$\|\hat{\boldsymbol{\theta}} - \boldsymbol{\theta}_0\| \xrightarrow{\mathcal{P}} 0.$$

Having stated the consistency, we can now give the conditions for the asymptotic normality of $\hat{\boldsymbol{\theta}}$. Defining $N_J$ as the smallest number of wavelet coefficients among all the scales of decomposition, these conditions are as follows:

**(C7)** $\boldsymbol{\theta}_0$ is an interior point to $\boldsymbol{\Theta}$.
**(C8)** $\mathbf{H}(\boldsymbol{\theta}_0) \equiv \frac{\partial}{\partial \boldsymbol{\theta} \partial \boldsymbol{\theta}^T}\boldsymbol{\nu}(\boldsymbol{\theta})\big|_{\boldsymbol{\theta}=\boldsymbol{\theta}_0}$ exists and is non-singular.
**(C9)** $\sqrt{N_J}\boldsymbol{s}^T\mathbf{V}^{-1/2}(\hat{\boldsymbol{\nu}} - \boldsymbol{\nu}) \xmapsto[N_J \to \infty]{\mathcal{D}} \mathcal{N}(0,1)$, where $\|\boldsymbol{s}\| = 1$.

Condition **(C7)** is also a standard condition since normality is commonly proven via expansions based on derivatives which cannot be made if the true parameter is at the bounds of the parameter space $\boldsymbol{\Theta}$. Condition **(C8)** is also usually assumed since it depends on the specific model $F_{\boldsymbol{\theta}}$ from which the random field $(X_k)$ is generated and cannot therefore be proven in general. On the other hand, Condition **(C9)** is also essential for the asymptotic normality of $\hat{\boldsymbol{\theta}}$ and it has been shown that $\hat{\boldsymbol{\nu}}$ verifies this condition under the specific assumptions highlighted in Guerrier and Molinari (2016b). In the latter, the conditions on mean-square consistency and joint asymptotic normality of the estimator $\hat{\boldsymbol{\nu}}$ are given for the first time, thereby allowing to understand the conditions needed for the asymptotic properties of the RGMWM to hold. More specifically, aside from the conditions needed for the consistency of $\hat{\boldsymbol{\nu}}$ mentioned earlier, the additional conditions needed for **(C9)** to be verified are that the WV score function is twice Bouligand differentiable and that, for example, $J < \lfloor \log_2(\sqrt{N}) \rfloor$ when using the Haar wavelet filter in a time series setting, where $\lfloor x \rfloor$ represents the largest integer smaller than $x$. Having discussed these conditions, we can use them to state the following proposition.

**Proposition 2.2.** *Under the conditions for Lemma 2.1 as well as Conditions (C7) to (C9), the estimator $\hat{\boldsymbol{\theta}}$ has the following asymptotic distribution*

$$\sqrt{N_J}\left(\hat{\boldsymbol{\theta}} - \boldsymbol{\theta}_0\right) \xrightarrow[N_J \to \infty]{\mathcal{D}} \mathcal{N}\left(\mathbf{0}, \mathbf{BVB}^T\right)$$

*where* $\mathbf{B} = \mathbf{H}(\boldsymbol{\theta}_0)^{-1}\mathbf{D}(\boldsymbol{\theta}_0)^T \boldsymbol{\Omega}$ *and* $\mathbf{D}(\boldsymbol{\theta}_0) = \partial/\partial\boldsymbol{\theta}\,\boldsymbol{\nu}(\boldsymbol{\theta})|_{\boldsymbol{\theta}=\boldsymbol{\theta}_0}$.

The proofs of the above propositions can be found in Appendix B. The results of Guerrier and Molinari (2016b) are therefore essential to obtain the asymptotic properties of the RGMWM estimator stated in Propositions 2.1 and 2.2. Moreover, following Genton and Ronchetti (2003), choosing a bounded $\psi$-function for the M-estimator in Guerrier and Molinari (2016b) ensures robustness of $\hat{\boldsymbol{\nu}}$ allowing the RGMWM estimator $\hat{\boldsymbol{\theta}}$ to be robust due to the bounded IF of $\hat{\boldsymbol{\nu}}$. This new estimator delivers a few theoretical advantages over existing robust estimators as well as considerable practical advantages as shown in Section 3. First of all, it can deal with non-Gaussian processes as emphasized in Guerrier and Molinari (2016b) and benefits from different parameter identifiability results given in Guerrier and Molinari (2016a). Moreover, the conditions for joint asymptotic normality are known for mulitdimensional random fields and can therefore be taken into account for the asymptotic properties of the RGMWM to hold. A main advantage however resides in the fact that the dimension of the auxiliary parameter vector is always reasonable since in general $J \leq \lfloor \log_2(N) \rfloor - 1$ which allows to make use of all the scales of WV without the need to select specific moments even for extremely large sample sizes. This is not the case, for example, for GMM estimators where moment-selection is an important issue since, according to the model that is being estimated, the choice should fall on all moments (which can be highly impractical) or on moments that are more "informative" than others (Andrews, 1999). The RGMWM on the other hand makes use of the WV which adequately summarizes all the information in the spectral density into a few auxiliary parameters without the need to select specific moments which contain more information.

Considering the different conditions listed above, let us state the following corollary which gives the conditions for the asymptotic properties of a specific RGMWM estimator. This estimator will then be used in the simulations and applications of the following sections since it is reasonable to assume that it can be commonly used in practice.

**Corollary 2.1.** *Assuming that $(X_k)$ is a Gaussian random field, that $\boldsymbol{\nu}$ belongs to a compact set and that $\mathbf{S}_\psi(\mathbf{0}) > 0$, under Conditions (C1), (C3) and Conditions (C6) to (C8), the RGMWM based on the Tukey $\psi$-function is consistent and asymptotically normally distributed considering $J$ as fixed.*

The proof simply follows from the results given in Guerrier and Molinari (2016b) and in this section. Indeed, most of the conditions in Corollary 2.1 are parameter- and/or model-specific and therefore have to be assumed in general or proven case-by-case. Moreover, under the Gaussian assumption for $(X_k)$, we know that $\boldsymbol{\nu}$ is identifiable and that Condition (C4) is verified using the Tukey $\psi$-function (see Guerrier and Molinari, 2016b). Having defined this new robust estimator, the following sections investigate its performances over a series of simulations studies and applications for time series and spatial model estimation.

## 3. Simulation Studies

The aim of this section is to show that the RGMWM estimator has a reasonable performance in settings where there is no contamination and has a better performance than the classical (and possibly robust) alternatives when the data are contaminated. Concerning the robust alternatives, as explained in further details in the following sections, there is a lack of implemented robust methods for time series analysis and complete absence for the estimation of spatial models. For this reason, we only make comparisons with methods that we were able to obtain good results with. Moreover, we intend to show the numerical stability and computational efficiency of this estimator, especially in the time series setting with models of higher complexity than an AR($p$) model.

To measure the statistical performance of the estimators we choose to use a robust and relative version of the Root Mean Squared Error (RMSE) defined as follows

$$\text{RMSE}^* = \sqrt{\text{med}\left(\frac{\hat{\theta}_i - \theta_{i,0}}{\theta_{i,0}}\right)^2 + \text{mad}\left(\frac{\hat{\theta}_i}{\theta_{i,0}}\right)^2}$$

with med($\cdot$) representing the median, mad($\cdot$) the median absolute deviation and $\hat{\theta}_i$ and $\theta_{i,0}$ representing the $i^{th}$ element of the estimated and true parameter vectors respectively. The RMSE* is therefore related to the RMSE and can also be used to assess the accuracy of an estimator. The classical RMSE was also used allowing to reach equivalent conclusions but the RMSE* was preferred to better highlight the difference between methods.

### *3.1. Time Series Model Estimation*

For the simulation studies on the estimation of time series models, 500 samples of size 1000 were generated for each type of model described further on. Different types of contamination were used to study the RGMWM, going from scale-contamination to additive and replacement outliers as well as patchy outliers and level-shifts. Innovation-type contamination was not considered since it did not appear to affect the estimators much (see Maronna et al., 2006, for an overview of different contamination settings). We denote the proportion of contaminated observations with $\epsilon$ and the size of contamination (i.e. the variance of the observations which are added to the uncontaminated observations) with $\sigma_\epsilon^2$. Finally, when dealing with level-shifts, we denote $\mu_{\epsilon_i}$ as the size of the $i^{th}$ shift in level.

Although the RGMWM is mainly conceived for the robust estimation of *latent* time series models, to compare it with other classic and robust estimators we choose to study its behaviour mainly on standard ARMA models for which it is also a consistent estimator based on the conditions in Guerrier and Molinari (2016a). In this perspective, we compare the proposed RGMWM estimator which is implemented in the `gmwm` R package (see Balamuta et al., 2016) with:

- the Maximum Likelihood estimator (ML);
- the M-estimator for autoregressive models proposed by Kunsch (1984) (MAR);
- the Indirect Inference estimator based on the MAR (INDI) (see de Luna and Genton, 2001);
- the standard GMWM estimator (GMWM);
- the GMWM estimator used in Guerrier et al. (2014) (MPWM).

Given the substantial absence of general routines in statistical software to robustly estimate time series models, the choice of alternative methods available for comparison was reduced. Among the many methods proposed, most of them required the computation of the explicit form of the score function or of a specific link function for each model and, once this is done, the tuning of certain parameters such as specific weighting matrices or others. Moreover, if these steps are achieved, the numerical stability of these methods is not necessarily guaranteed requiring an additional step to fine-tune the implementation. Given this and failing to obtain readily usable code from different authors, the choice of the methods was finally based on their direct availability within statistical software or their easy implementation based on these available tools. This corresponds to the setting in which a general researcher or practitioner, with basic statistical knowledge, would like to estimate model parameters in a robust manner. In this logic, using the robust regression tools available in the statistical software R, we implemented the robust MAR estimator proposed by Kunsch (1984) and, based on this, the INDI represents the corresponding easy-to-implement estimator for ARMA models (other more computationally efficient methods have been proposed such as the method proposed in Ortelli and Trojani, 2005). Nevertheless, when the latter estimator was used, the number of simulations for indirect in-ference was set to $H = 30$ since otherwise the computational time was beyond 5 minutes for each estimation and this issue was not improved by modifying the order of the auxiliary AR($p$) model. The MAR estimator was used exclusively for the estimation of AR($p$) models while the INDI estimator was used in all the other settings and was unweighted (i.e. the weighting matrix for indirect inference was the identity matrix). Considering this last aspect, to make the comparison fair, the GMWM, RGMWM and MPWM also used an unweighted matrix when being compared to the results of the INDI estimator. Moreover, considering the cases where the INDI estimator was used, a preliminary simulation study was carried out to determine the order of the auxiliary AR($p$) model. As for the starting values, the INDI and MPWM used the ML estimates as starting values while the GMWM and RGMWM also use the ML within the starting-value algorithm of the `gmwm` package in the `R` statistical software. Finally, considering the tuning parameters of the robust estimators, the MPWM estimator uses the median-type WV estimator used for the simulation study in Mondal and Percival (2012b) and therefore is highly robust. For this reason, the level of efficiency chosen for the MAR and RGMWM estimators was of 0.6 in order to guarantee a comparable level of robustness. Since the RGMWM is based on the WV estimated through the Tukey biweight function, this function was chosen also for the MAR estimator and, after a preliminary simulation study to determine the level of efficiency, a tuning constant $c_{MAR} = 2.2$ was chosen.

The performance of these estimators is investigated on the following models and contamination settings:

- **AR(1)**: a zero-mean first-order autoregressive model with parameter vector
  $[\rho_1 \ \upsilon^2]^T = [0.9 \ 1]^T$, scale-based contamination at scale $\jmath = 3$, $\epsilon = 0.01$ and $\sigma_\epsilon^2 = 100$;
- **AR(2)**: a zero-mean second-order autoregressive model with parameter vector $[\rho_1 \ \rho_2 \ \upsilon^2]^T = [0.5 \ -0.3 \ 1]^T$, isolated outliers, $\epsilon = 0.05$ and $\sigma_\epsilon^2 = 9$;
- **ARMA(1,2)**: a zero-mean autoregressive-moving average model with parameter vector $[\rho \ \varrho_1 \ \varrho_2 \ \upsilon^2]^T = [0.5 \ -0.1 \ 0.5 \ 1]^T$, and level-shift contamination with $\epsilon = 0.05$, $\mu_{\epsilon_1} = 5$ and $\mu_{\epsilon_2} = -3$;
- **ARMA(3,1)**: a zero-mean autoregressive-moving average model with parameter vector $[\rho_1 \ \rho_2 \ \rho_3 \ \varrho_1 \ \upsilon^2]^T = [0.7 \ 0.3 \ -0.2 \ 0.5 \ 2]^T$, patchy outliers, $\epsilon = 0.01$ and $\sigma_\epsilon^2 = 100$;
- **SSM**: a state-space model ($X_t$) interpreted as a composite (latent) process in certain engineering applications. This model is defined as

$$Y_t^{(i)} = \rho_{(i)} Y_{t-1}^{(i)} + W_t^{(i)}$$
$$W_t^{(i)} \stackrel{\text{iid}}{\sim} \mathcal{N}(0, \upsilon_{(i)}^2)$$
$$X_t = \sum_{i=1}^{2} Y_t^{(i)} + Z_t,$$
$$Z_t \stackrel{\text{iid}}{\sim} \mathcal{N}(0, \sigma^2)$$

with parameter vector

$$[\rho_{(1)}\ v^2_{(1)}\ \rho_{(2)}\ v^2_{(2)}\ \sigma^2]^T = [0.99\ 0.1\ 0.6\ 2\ 3]^T,$$

isolated outliers, $\epsilon = 0.05$ and $\sigma^2_\epsilon = 9$.

For each simulation, the sample size is $T = 1,000$ which delivers $J = 9$ scales for the GMWM-type estimators. This is one possible limitation of the latter estimators since, for identifiability issues, one needs at least as many moments or auxiliary parameters as the number of parameters of interest. Given the sample size, the larger models will rely more on the larger scales for which the WV estimators are less efficient and we therefore expect a decreased performance of these estimators for these models given this sample size. The ML and INDI estimators were not considered for the **SSM** simulations since they always failed to converge or gave questionable results using available software such as the `MARSS` package in the `R` statistical software. A smaller simulation study with $T = 100$ was carried out for this model and even in this case the convergence rates for these estimators were below 30%. This is highlighted in Table 1 which reports summary information regarding the estimation time in seconds for the RGMWM and INDI estimators in the contaminated setting for the **ARMA(1,2)**, **ARMA(3,1)** and **SSM** models along with their convergence rates (the results for the uncontaminated setting can be found in Appendix C). Having stated this, the results of the simulation studies using the RMSE* are shown in Figure 1 where the MAR and INDI estimators are denoted as KUN-SCH since they are used in complementary settings.

FIG 1. *Top row: RMSE\* of the estimators in an uncontaminated setting. Bottom row: RMSE\* of the estimators in a contaminated setting. KUNSCH represents the MAR estimator for the **AR(1)** and **AR(2)** models while it represents the INDI estimator for the other models.*

|  | Model | Sample size | Median | Conv. rate (%) |
|---|---|---|---|---|
| RGMWM | **ARMA(1,2)** | 1,000 | $5.59 \cdot 10^{-1}$ | 100 |
|  | **ARMA(3,1)** | 1,000 | $7.9 \cdot 10^{-1}$ | 100 |
|  | **SSM** | 100 | $5.6 \cdot 10^{-2}$ | 100 |
| INDI | **ARMA(1,2)** | 1,000 | $1.79 \cdot 10^{2}$ | 70 |
|  | **ARMA(3,1)** | 1,000 | $2.094 \cdot 10^{2}$ | 92 |
|  | **SSM** | 100 | $1.025 \cdot 10^{2}$ | 29.2 |

TABLE 1
*Sample size, median computational time in seconds and convergence rates of the RGMWM and INDI estimators for the models **ARMA(1,2)**, **ARMA(3,1)** and **SSM** in a contaminated setting.*

The first aspect to underline is that the RGMWM generally performs better than the MPWM in all the different settings and can therefore be considered as an improvement over the robust estimator investigated in Guerrier et al. (2014). Having stated this, when considering the **AR(1)**, **AR(2)** models, the RGMWM does not lose much in uncontaminated settings while it performs generally as well or better than the MAR estimator in contaminated ones. As for the **ARMA(1,2)** model, the RGMWM is not as efficient as the others (excluding the MPWM) in the uncontaminated case while it adequately bounds the influence of outliers in contaminated ones, performing roughly as well or better than the INDI estimator. As for the **ARMA(3,1)** model, the RGMWM performs as well as the other estimators in uncontaminated settings while it clearly is the best estimator overall in contaminated ones. This can be related to the ease with which the GMWM estimators can estimate more complex models from a numerical point of view. The latter is confirmed looking at the results for the **SSM** simulations where it can be seen how the RGMWM is extremely close to the GMWM in uncontaminated settings while it is overall the best estimator in the contaminated ones. This is a clear advantage of the RGMWM since it provides a computationally efficient and numerically stable method to robustly estimate the parameters of many linear state-space models which has been almost unfeasible in practice to date.

There are a few points that must be emphasized considering the above results. First of all, as Table 1 shows, the RGMWM can estimate all these models in under a second while the INDI estimator can take up to 300 times more, with timings going up to more than 8 minutes, and does not consider the timing for those cases in which it does not converge numerically. Secondly, the INDI estimator can eventually use the RGMWM as a starting value thereby avoiding to use an algorithm to find adequate starting values which can increase its computational times even more. Moreover, it must be emphasized again that the RGMWM (as the other GMWM estimators) does not use a full weighting matrix in these cases and consequently its performance could greatly be improved as seen in the **AR(1)** and **AR(2)** model simulations.

*3.2. Spatial Model Estimation*

In this section we study the RGMWM in the estimation of spatial models. For this purpose, regular lattice $K \times M$ random fields were simulated 500 times, with $K = M = 30$, thereby delivering sample sizes of $N = 900$ and $J = 10$ scales of wavelet decomposition since we limit ourselves to isotropic models which therefore only require the lower triangular WV matrix. Larger sample sizes were not considered since this became computationally impractical for the maximum likelihood estimator which is already time-demanding for this problem size. In this case, the following estimators were considered in addition to the RGMWM:

- the Maximum Likelihood estimator (ML);
- the standard GMWM estimator (GMWM);
- the GMWM estimator based on the robust estimator of WV proposed by Mondal and Percival (2012b) (MPWM).

To the best of our knowledge, there is no existing proposed or implemented robust procedure for the estimation of the standard spatial models considered in this section. Therefore the comparison is only made with the ML and GMWM that are non-robust estimators.

The performance of these estimators is investigated on the following models and contamination settings:

- **Exp(1)**: a zero-mean Exponential model with parameter vector $[\phi \; \sigma^2]^T = [2 \; 1]^T$ and level-shift contamination with $\epsilon = 0.05$, $\mu_{\epsilon_1} = 5$ and $\mu_{\epsilon_2} = -3$;
- **Exp(2)**: a sum of two zero-mean Exponential models with parameter vector 
  $[\phi_{(1)} \; \sigma^2_{(1)} \; \phi_{(2)} \; \sigma^2_{(2)}]^T = [2 \; 1 \; 1.5 \; 1]^T$, isolated outliers, $\epsilon = 0.01$ and $\sigma^2_\epsilon = 100$;
- **Gauss(1)**: a zero-mean Gaussian model with parameter vector $[\phi \; \sigma^2]^T = [2 \; 1]^T$, patchy outliers, $\epsilon = 0.01$ and $\sigma^2_\epsilon = 100$;
- **Gauss(2)**: a sum of two zero-mean Gaussian models with parameter vector 
  $[\phi_{(1)} \; \sigma^2_{(1)} \; \phi_{(2)} \; \sigma^2_{(2)}]^T = [2 \; 1 \; 1.5 \; 1]^T$, isolated outliers, $\epsilon = 0.05$ and $\sigma^2_\epsilon = 9$;

The only model for which the ML was used was the **Exp(1)** model since it was unable to estimate the Gaussian and the latent models due to numerical instability using either the `likfit()` function in the R package `geoR` or a tailored implementation of the ML. With this in mind, similarly to Section 3.1, the RMSE* of these estimators for the above models are given in Figure 2.

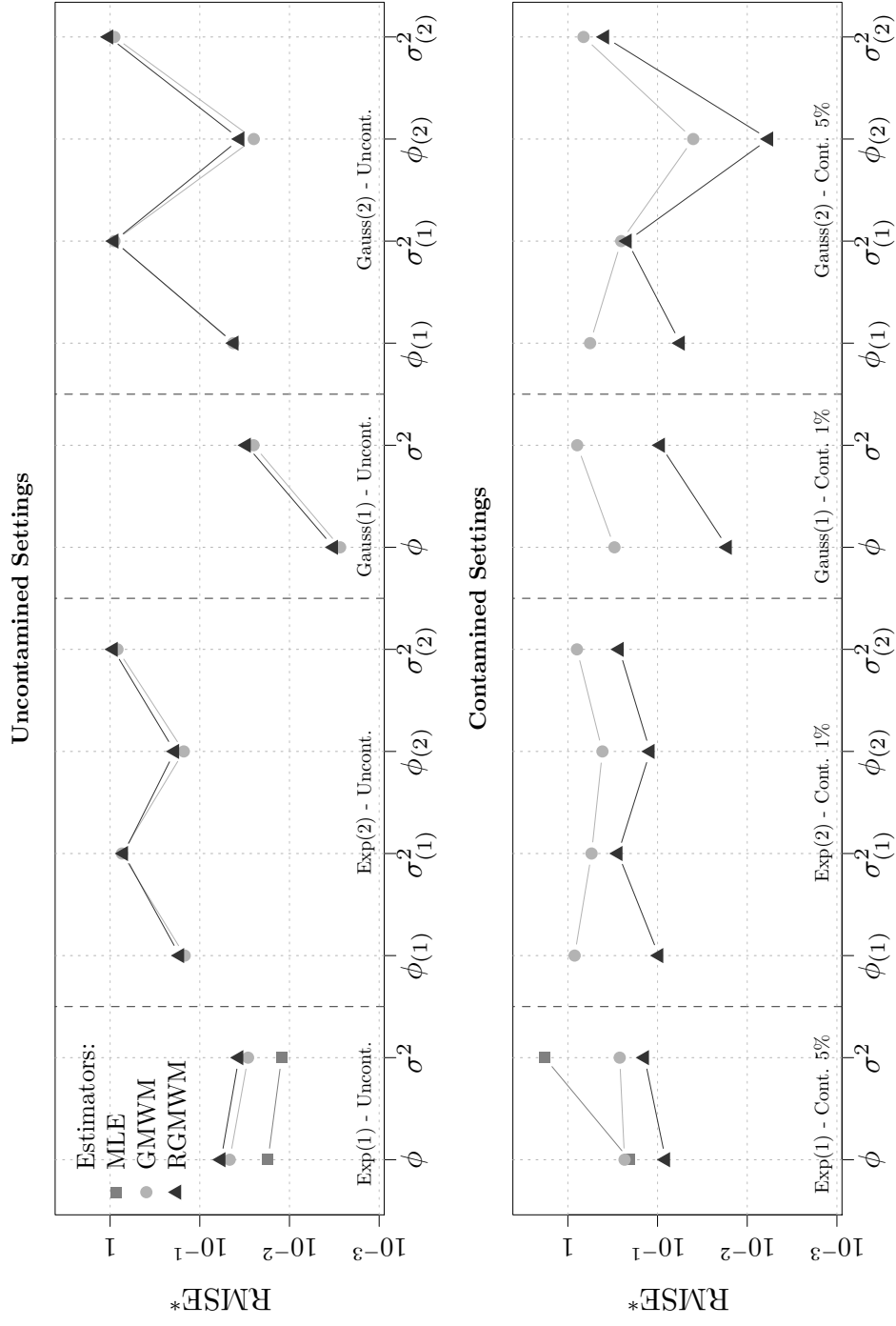

FIG 2. *Top row: RMSE\* of the estimators in an uncontaminated setting. Bottom row: RMSE\* of the estimators in a contaminated setting.*

The first aspect to underline is that, in any case, the GMWM and RGMWM estimators provide computationally efficient and statistically sound solutions for the estimation of standard and latent spatial models which are often numerically challenging for existing methods such as the ML (which could only be used for the **Exp(1)** model). Nevertheless, focusing on the simulation results for the **Exp(1)** model, the ML is by far the best estimator in the uncontaminated setting and we would expect similar results for the **Gauss(1)** model (and the others) if there weren't any numerical issues. However, it is also clear that performance of the ML is worse under contamination where the RGMWM, being the only robust estimator available, is the best in bounding the influence of outliers. In all other settings, we can see that the GMWM is slighlty better than the RGMWM in the uncontaminated settings whereas the performance of the latter does not change much under contamination and therefore makes it the preferred estimator in these settings.

To summarize, as the simulation studies in this section and in Section 3.1 have shown, the RGMWM estimator represents a good robust and general purpose estimator for time series and spatial models, even when these are of considerable complexity and the number of observations is large, delivering estimation and inference results in a computationally efficient and numerically stable manner.

## 4. Applications

The wide class of intrinsically stationary models that the RGMWM can estimate in a robust manner allows it to be used in a large variety of applications where outliers and contamination can often occur. In this section we will therefore investigate the performance of the proposed RGMWM estimator on some real data sets coming from:

- Engineering;
- Economics;
- Hydrology;
- Climatology.

In all these applications different models are used, going from simple AR(1) models (for time series) and Exponential models (for spatial data) to complex latent models which include ARMA models. To select these models, a bootstrap version of the J-test (see Hansen, 1982) was used, as suggested in Guerrier et al. (2013), therefore selecting the models for which we could not reject the null hypothesis that they fit the data well.

### 4.1. Application to Inertial Sensors

The engineering data set consists in the angular rate signal issued from a micro-electro-mechanical system gyroscope in static conditions. Due also to their low cost, these sensors are very common and are being increasingly used in the field of navigation engineering. The main goal of recording this kind of data is to improve the performance of the navigation sensors by identifying and estimating the parameters of the error model coming from the accelerometers and gyroscopes that compose the sensor. Once these parameters are estimated they are inserted in a filter (usually an extended Kalman filter) which is used within a navigation system. The latter collects measurements from different sources such as Global Positioning Systems (GPS) or the inertial sensors themselves in an optimal manner in order to improve the navigation precision. Therefore, the latter greatly depends on the estimation of the parameters of the selected error model for the inertial sensor.

Figure 3 shows the error signal from the gyroscope along with the outliers in a portion of the signal identified via the weights given to the observations by the RGMWM estimator. As can be observed, there are outliers that would appear to be obvious by simply looking at the plot and could be treated by fault detection algorithms for navigation systems (see further on) but there are many others that lie within the part of the signal which one would not expect to contain outliers. Despite the numerous outliers, these are extremely low in proportion to the whole dataset ($\approx 0.4\%$) which contains a little under 900,000 observations (issued from an approximately 2.5 hours-long recording sampled at 100 Hz). This may lead to think that estimations on this dataset would not be significantly influenced by outliers.

Nevertheless, to understand how influential these observations could be, we estimated the classical and robust WV from the signal represented in Figure 3. Using these estimates we then estimated an error model made by the sum of three latent first-order autoregressive models. This state-space model is among those suggested by Stebler et al. (2014) as being most appropriate to describe such signals. Table 7 in Appendix D shows the estimated parameters for the GMWM and RGMWM estimators together with their confidence intervals (the ML was not considered for the same reasons given in Section 3 for the **SSM** model: the numerical stability and computational efficiency are unreasonable for this model and sample size). For both estimators the values of some autoregressive parameters are close to one, suggesting that the AR(1) model could be considered as a random walk. Indeed, a model that was commonly used to describe these signals was the sum of a white noise process with a random walk. However, Stebler et al. (2014) show how the use of sums of AR(1) models greatly improves the navigation performance over this model and the J-tests and confidence intervals support this view by ruling out the models which included a random walk. Although the differences between the estimations do not appear to be large since the estimated level of contamination is low, a significant difference is to be noticed for the parameters of the first two autoregressive processes indicating that the contamination appears to have an impact on estimation and that robust methods should be preferred (assuming the Gaussian assumption holds). Even one (or few) slightly misestimated parameter(s) can be highly relevant in the context of navigation systems since these are fed into the filters which will progressively misestimate the position as the sensors work in "coasting mode" (i.e. without the GPS integration) and deliver the so-called "error accumula-tion". Informally speaking this is due to the fact that these measurements are integrated several times and therefore their errors accumulate in time especially when no GPS observations are present to "reinitialize" the system (more details on this can be found, for example, in Titterton and Weston, 2004).

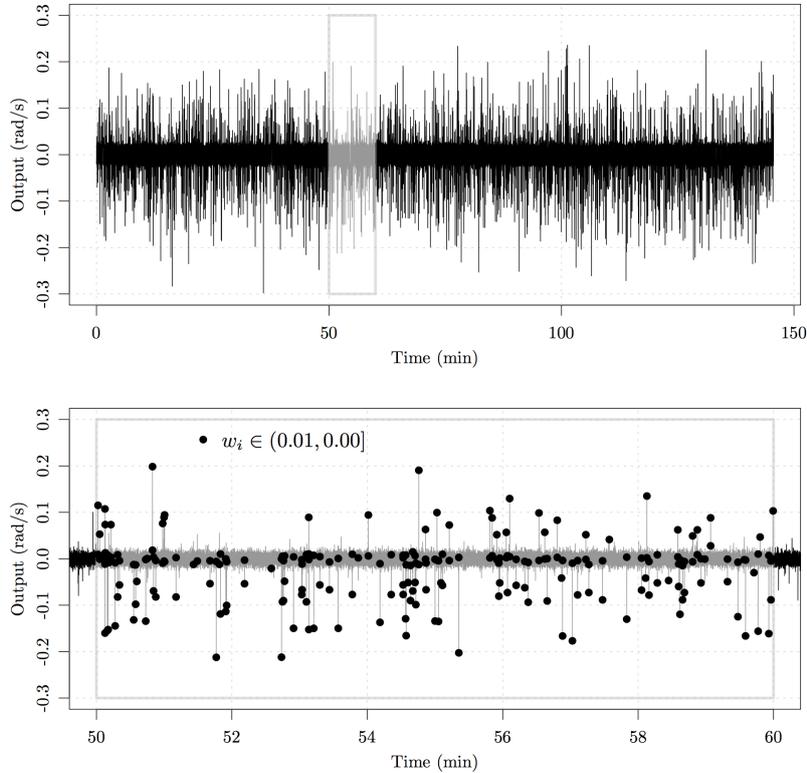

FIG 3. *Top part: Inertial sensor time series. Bottom part: zoom-in on grey part of the time series with black points indicating extreme outliers identified through the weights of RGMWM.*

Moreover, our robust approach can be of great usefulness in the area of Fault Detection and Isolation (FDI) for inertial measurement units (see for example Guerrier et al., 2012, and references therein) as shown in Figure 3. In general, the task of FDI includes the detection of the presence of failures (or outliers) and the isolation of the component responsible of the irregularity. In the inertial navigation framework, FDI algorithms are used, for example, to ensure the safety of aircrafts or robots which deeply rely on inertial sensors. In fact, usual FDI methods in this area use various measurements coming from several sensors which entail a series of disadvantages. Moreover, these methods often make use of unconditional cut-off values which generally determine which observations are "unusual". On the other hand, the proposed approach would be able to detect "unusual" observations conditioned on previous ones instead of detecting them based on a general cut-off value. Although this is left for future research as some further adjustments would need to be put in place, our approach could be used as a basis for FDI by only using one signal coming from the sensor calibration procedure. One of the advantages of this approach, in addition to those already mentioned, is that it would have important impacts in terms of costs and constraints (e.g. weight, electric consumption, etc.) for robots or small unmanned aerial vehicles which are currently a major focus of technological and mechanical research.

### 4.2. Application to Personal Saving Rates

The economics data set consists in the monthly seasonally adjusted Personal Saving Rates (PSR) data from January 1959 to May 2015 provided by the Federal Reserve Bank of St. Louis. The study of PSR is an essential part of the overall investigation on the health of national and international economies since, within more general economic models, PSR can greatly impact the funds available for investment which in turn determine the productive capacity of an economy. Understanding the behaviour of PSR is therefore an important step in correct economic policy decision making. In this sense, Slacalek and Sommer (2012) study the factors behind saving rates and investigate different models which, among others, are compared to the random-walk-plus-noise (local level) model (RWN). As opposed to the latter model, various time-varying models are proposed in the literature to explain precautionary PSR together with risk aversion in the light of different factors such as financial shocks or others (see, for example, Videras and Wu, 2004; Brunnermeier and Nagel, 2008). Nevertheless, as emphasized in Pankratz (2012), modelling the time series with a stationary model, or a $d^{th}$-order non-stationary model such as an ARIMA, can be useful under many aspects such as, for example, to understand if a dynamic model is needed for forecasting and, if so, what kind of model is appropriate.

In this example, we consider the RWN model and, as in Section 4.1, we use the WV log-log plot and a J-test to understand what kind of model could fit the time series. By doing so, we find that a random walk plus an ARMA(2,1) fits the data well and therefore, in this case, we have that the "noise" in the RWN model is an ARMA(2,1). This can be seen in Figure 4 where, in the top part, the saving rate time series is represented along with the identified outliers and, in the bottom part, we see the log-log representation of the classic and robust estimated and model-implied WV respectively. Indeed, for the bottom part, the diagonal plots show the classic and robust estimations respectively, each with the estimated WV and the WV implied by the estimated model. The off-diagonal plots compare the classic and robust estimated WV (upper diagonal) and the WV implied by the GMWM and RGMWM model parameter estimates (lower diagonal). It can be seen how there is a significant difference between the classic and robust WV estimates, especially at the first scales where the confidence intervals of the estimated WV do not overlap (upper diagonal plot). This leads to a difference in the model-implied WV whose parameters have been estimated through the GMWM and RGMWM (lower diagonal plot).

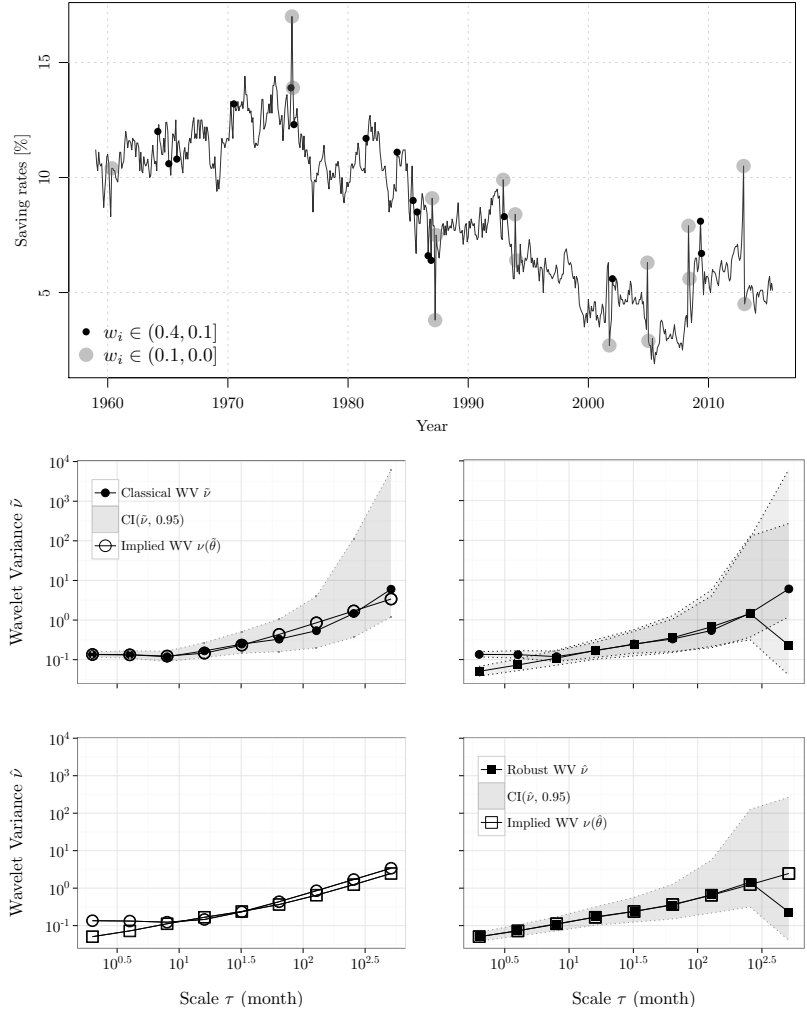

FIG 4. *Top figure: Saving rates time series with different types of points indicating outliers identified through the weights of the RGMWM. Bottom figure: log-log scale WV plots for saving rates series; Top left: classic estimated WV superposed with model-implied WV based on the parameters estimated through the GMWM. Top right: classic and robust estimated WV with respective confidence intervals superposed. Bottom left: classic and robust model-implied WV based on the GMWM and RGMWM estimates respectively. Bottom right: robust estimated WV superposed with model-implied WV based on the parameters estimated through the RGMWM.*

Table 2

*Random Walk plus ARMA(2,1) model estimates for the PSR data. Estimated parameters with GMWM and RGMWM estimators with $\gamma^2$ being the random walk parameter, $\rho_i$ the $i^{th}$ autoregressive parameter, $\varrho$ the moving average parameter and $\sigma^2$ the innovation variance of the ARMA(2,1) model. Confidence intervals (CI) based on the approach used in Guerrier et al. (2013).*

|  | GMWM | | RGMWM | |
|---|---|---|---|---|
|  | Estimate | CI($\cdot$, 95%) | Estimate | CI($\cdot$, 95%) |
| $\gamma^2$ | $7.95 \cdot 10^{-2}$ | ($3.67 \cdot 10^{-2}$ ; $1.11 \cdot 10^{-1}$) | $5.85 \cdot 10^{-2}$ | ($1.54 \cdot 10^{-2}$ ; $9.97 \cdot 10^{-2}$) |
| $\rho_1$ | $1.64 \cdot 10^{-1}$ | ($5.93 \cdot 10^{-2}$ ; $2.89 \cdot 10^{-1}$) | $6.00 \cdot 10^{-1}$ | ($4.48 \cdot 10^{-1}$ ; $7.55 \cdot 10^{-1}$) |
| $\rho_2$ | $3.06 \cdot 10^{-3}$ | ($-1.31 \cdot 10^{-1}$ ; $1.48 \cdot 10^{-1}$) | $1.84 \cdot 10^{-1}$ | ($3.10 \cdot 10^{-2}$ ; $2.46 \cdot 10^{-1}$) |
| $\varrho$ | $2.43 \cdot 10^{-1}$ | ($2.02 \cdot 10^{-1}$ ; $2.81 \cdot 10^{-1}$) | $2.92 \cdot 10^{-1}$ | ($2.28 \cdot 10^{-1}$ ; $3.45 \cdot 10^{-1}$) |
| $\sigma^2$ | $3.14 \cdot 10^{-1}$ | ($2.59 \cdot 10^{-1}$ ; $3.85 \cdot 10^{-1}$) | $1.32 \cdot 10^{-1}$ | ($8.59 \cdot 10^{-2}$ ; $1.80 \cdot 10^{-1}$) |

The estimated parameters with the GMWM and RGMWM are given in Table 2 along with their respective confidence intervals. There are two main differences between the two estimations: (i) the estimates of the first autoregressive parameter $\rho_1$ and innovation variance $\tilde{\sigma}^2$ are significantly different; (ii) the second autoregressive parameter $\rho_2$ is not significant using the GMWM. These differences highlight how the conclusions concerning parameter values and model selection can considerably change when outliers are present in the data. Indeed, the choice of the model would then affect the decisions taken towards the selection of appropriate causal and dynamic models to better explain the behavoiur of saving rates and the economy as a whole. The selected model based on the robust fit can in fact be interpreted as a sum of latent models along the lines given in Slacalek and Sommer (2012) where the ARMA(2,1) can be seen as a sum of two AR(1) models where each of them represents, for example, the reaction of PSR to changes in uncertainty (affected by unemployment) and interest rates, respectively, while the random walk describes the continuous fluctuations of target wealth which also drives PSR.

The additional benefit of the RGMWM estimator, as opposed to the median-type MPWM, is also to deliver weights that allow to identify outliers which may not be visible simply by looking at the time series (as highlighted in Section 4.1). As shown in the top part of Figure 4, the outliers identified by the RGMWM can be interpreted in the light of the national and global economic and political events. Limiting ourselves to the major identified outliers, the first one corresponds to a rise in the precautionary savings in the aftermath of the OPEC oil crisis and the 1974 stock market crash. In the months following October 1987 we can see an instability in the PSR with a rise and sudden fall linked to the "Black Monday" stock market crash which added to the savings and loans crisis which lasted to the early 1990s. This period also saw an economic recession where a rise in the saving rates, highlighted by the presence of high outliers, led to a drop in aggregate demand and bankruptcies. Finally, the various financial crises of the $21^{st}$ century led to sudden and isolated rises in PSR as indicated again by the outliers.

## *4.3. Application to Precipitation Data*

The hydrology data set collects the monthly precipitation data from 1907 to 1972 provided in Hipel and McLeod (1994) and is shown in the top panel of Figure 5. The modelling approach described in this section is the Environmental System Model (ESM) of a watershed which, despite being less used due to other more recent approaches (such as, for example, adaptive neural networks in Tokar and Johnson, 1999), can still be highly useful for practitioners who wish to have a straightforward and clear tool to describe and interpret phenomena linked to the water cycle. Moreover, the example clearly shows how our method can help detect dependence where classical methods may not due to contamination in the data (be this in the domain of hydrology or others). The goal of the ESM is to explain how water resources behave and are distributed throughout their cycles from the stage of precipitation to river flows. Salas and Smith (1981) describe how the precipitation model is the basis for the models of the following stages in the ESM. Three models are envisaged by Salas and Smith (1981) for the precipitation stage among which the independent precipitation (i.e. a white noise process) and the AR(1) model.

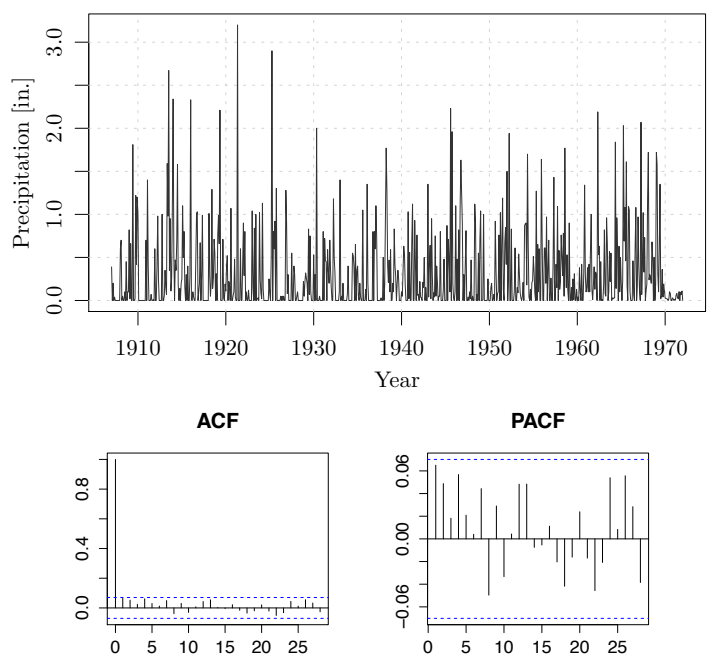

FIG 5. *(Top panel) Monthly precipitation series from 1907 to 1972 taken from Hipel and McLeod (1994). (Bottom panels) Estimated autocorrelation function (left) and estimated partial autocorrelation function (right) of the precipitation series.*

Taking a look at the time series, it may not appear to be Gaussian but the GMWM and RGMWM are based on the inherent first-differencing of the series through the Haar wavelet filter so they can be used in this case. Analyzing the AutoCorrelation Function (ACF) and Partial AutoCorrelation Function (PACF) in the bottom panels of Figure 5 one would identify an independent precipitation model for the this dataset. However, an AR(1) model was fitted to understand if the independent model was reasonable.

Table 3 shows the estimated parameters for the AR(1) precipitation model. The latter model has different estimates between the classical estimators and the proposed robust one. In fact, the ML and GMWM estimates tend to agree with the independent model assumption where the Confidence Intervals (CI) for the autoregressive parameter are close to or include the value of zero, whereas the RGMWM detects a stronger dependence with the previous precipitation measurement and a smaller variance of the innovation process (with CI not overlapping those of the ML and GMWM). This could be due to the fact that the classical ACF and PACF are sensitive to outliers and may not detect this correlation structure (see Maronna et al., 2006).

TABLE 3
*AR(1) estimates for the mean monthly precipitation data from 1907 to 1972 taken from Hipel and McLeod (1994). Estimated parameters with ML, GMWM and RGMWM estimators with $\hat{\phi}$ being the estimated autoregressive parameter and $\hat{\sigma}^2$ the innovation variance. Confidence Intervals (CI) based on the approach used in Guerrier et al. (2013).*

|       | $\hat{\phi}$ | $\hat{\sigma}^2$ |
|-------|---|---|
| ML    | $6.843 \cdot 10^{-2}$ | $2.199 \cdot 10^{-1}$ |
| CI    | $[1.110 \cdot 10^{-3}, 1.212 \cdot 10^{-1}]$ | $[2.003 \cdot 10^{-1}, 2.3929 \cdot 10^{-1}]$ |
| GMWM  | $5.577 \cdot 10^{-2}$ | $2.179 \cdot 10^{-1}$ |
| CI    | $[-4.357 \cdot 10^{-3}, 1.153 \cdot 10^{-1}]$ | $[1.985 \cdot 10^{-1}, 2.365 \cdot 10^{-1}]$ |
| RGMWM | $4.049 \cdot 10^{-1}$ | $1.066 \cdot 10^{-1}$ |
| CI    | $[3.345 \cdot 10^{-1}, 4.662 \cdot 10^{-1}]$ | $[9.623 \cdot 10^{-2}, 1.175 \cdot 10^{-1}]$ |

If the ESM were to be used in this context, it would be greatly affected by a misspecified model for the precipitation since it would condition the model choice and relative parameter estimation in the following phases of the water cycle. In this example, the choice of an independent precipitation model would have lead to a domino-effect in terms of model misspecification and misestimation leading to possibly highly incorrect interpretations and conclusions.

## 4.4. Application to Cloud Data

The cloud data set consists in four regions selected from a temperature map of the south-east Pacific stratocumulus clouds near the Chilean coast made available by Geo-stationary Operational Environmental Satellites (GOES) Imagery (see Mondal and Percival, 2012b, for details on this data). These regions were selected by atmospheric scientists and have already been analysed in Mondal

|  | $\hat{\phi}$ | $\hat{\sigma}^2$ |
|---|---|---|
| GMWM | $1.35 \cdot 10^0$ | $8.443 \cdot 10^{-2}$ |
| CI | $[1.128 \cdot 10^0, 1.575 \cdot 10^0]$ | $[6.537 \cdot 10^{-2}, 1.035 \cdot 10^{-1}]$ |
| RGMWM | $1.5 \cdot 10^0$ | $2.415 \cdot 10^{-2}$ |
| CI | $[1.403 \cdot 10^0, 1.596 \cdot 10^0]$ | $[8.35 \cdot 10^{-3}, 3.994 \cdot 10^{-2}]$ |

TABLE 4

*Exponential model estimates for the cloud image data in the third region using the GMWM and RGMWM estimators. $\hat{\phi}$ is the estimated range parameter and $\hat{\sigma}^2$ the estimated sill. Confidence Intervals (CI) based on the approach used in Guerrier et al. (2013).*

and Percival (2012a) using the WV as a means to detect variance patterns according to the type of cloud formation. The study of these cloud formations and their related dynamics is of particular importance to understand the radiation patterns which this region is subject to, also considering the presence of specific industrial activities which release atmospheric aerosol and make this a complex region to study from a climatological point of view.

More specifically, Mondal and Percival (2012a) highlighted that the third and fourth regions delivered considerably different patterns when using the standard and robust estimators of WV. A possible reason for this difference can be identified in the fact that, as opposed to the first two regions, the third and the fourth depict inhomogeneous or varying scenarios since they represent respectively a formation of broken clouds and a forming Pocket of Open Cells (POC) which are clear areas in the clouds that are mainly characterized by low-aerosol air mass. Indeed, the first two regions represent a well-formed POC and a uniform formation of clouds respectively and, possibly due to this "stability", the variance patterns did not appear to change much between the standard and robust analysis. Correctly analysing the variance patterns is essential to identify the type of cloud formation and consequently select the appropriate model to describe these climatological phenomena. In Mondal and Percival (2012b) they consider an Exponential covariance model to describe these regions. Let us denote this covariance model as $\varphi(d)$, where $d$ is the Euclidean distance between two variables $X_k$ and $X_{k'}$ (with $k$ denoting the coordinates of a location), and therefore define the Exponential covariance model as

$$\varphi(d) = \sigma^2 \exp\left(-\frac{d}{\phi}\right),$$

where $\phi$ is the range parameter and $\sigma^2$ is the sill. After testing for isotropy using a robust version of the method proposed in Thon et al. (2015), we cannot reject the hypothesis that the images are isotropic, so we can therefore consider the Exponential model as a valid candidate. Using the GMWM and RGMWM to estimate its parameters, the results of these two methods on the third and fourth regions are presented in Tables 4 and 5 along with their confidence intervals.

For both regions, the estimates of the range parameter $\phi$ for the GMWM and RGMWM are not significantly different and substantially agree with the interpretation in Mondal and Percival (2012b) who, taking into account their

|  | $\hat{\phi}$ | $\hat{\sigma}^2$ |
|---|---|---|
| GMWM | $1.453 \cdot 10^0$ | $3.615 \cdot 10^{-2}$ |
| CI | $[1.385 \cdot 10^0, 1.52 \cdot 10^0]$ | $[2.914 \cdot 10^{-2}, 4.316 \cdot 10^{-2}]$ |
| RGMWM | $1.456 \cdot 10^0$ | $1.063 \cdot 10^{-2}$ |
| CI | $[1.416 \cdot 10^0, 1.496 \cdot 10^0]$ | $[4.225 \cdot 10^{-3}, 1.703 \cdot 10^{-2}]$ |

TABLE 5

*Exponential model estimates for the cloud image data in the fourth region using the GMWM and RGMWM estimators. $\hat{\phi}$ is the estimated range parameter and $\hat{\sigma}^2$ the estimated sill. Confidence Intervals (CI) based on the approach used in Guerrier et al. (2013).*

reparametrization, suggest a value of $\phi$ roughly between 1 and 2. However, a significant difference can be seen in the estimates of the sill parameter $\sigma^2$ which determines the unconditional variance of the random field. This parameter can make a considerable difference in practical situations where the estimated models are used as climatological models for simulation purposes in order to determine the evolution and interactions of cloud formations. An overestimation of the sill parameter $\sigma^2$ can introduce an undesirable increase in the variability of the simulation studies, thereby adding uncertainty to the conclusions made based on these models.

## 5. Conclusion

This paper presented a new framework for the robust estimation of time series and spatial models called RGMWM which extends to various classes of models that are intrinsically stationary. This framework provides estimators which are easy-to-implement, computationally efficient and have suitable asymptotic properties. The simulation studies and the applied examples confirm that the robust estimators delivered via the proposed approach adequately bound the influence of outliers on the estimation procedure and compare satisfactorily to alternative estimators which, with a few exceptions, are numerically challenging and/or computationally intensive. This paper hence provides a contribution in the direction of developing a general theoretical framework to robust inference for (latent) time series and spatial models as well as a method which is computationally efficient and straightforward to implement in practice thereby diminishing many of the challenges that researchers and practitioners are faced with when dealing with contamination in dependent data with possibly complex models and large sample sizes.

## References

C. Agostinelli and L. Bisaglia. ARFIMA processes and outliers: a weighted likelihood approach, 2010.

H. Allende and S. Heiler. Recursive generalized m estimates for autoregressive moving-average models. *Journal of Time Series Analysis*, 13:1–18, 1992.

Donald WK Andrews. Consistent moment selection procedures for generalized method of moments estimation. *Econometrica*, 67(3):543–563, 1999.

Luc Anselin. *Spatial econometrics: methods and models*, volume 4. Springer Science & Business Media, 2013.

J. Balamuta, S. Guerrier, R. Molinari, and W. Yang. The gmwm R package: a comprehensive tool for time series analysis from state-space models to robustness. *Submitted manuscript*, 2016.

A. M. Bianco, M. Garcia Ben, E. J. Martinez, and V. J. Yohai. Robust procedures for regression models with ARIMA errors. In *Proceedings in Computational Statistics, COMPSTAT 96*, pages 27–38, Heidelberg, 1996. Physica-Verlag.

A. G. Bruce, D. L. Donoho, H. Y. Gao, and R. D. Martin. Denoising and robust non-linear wavelet analysis. In *SPIE Proceedings, Wavelet Applications*, pages 325–336, 1994.

Markus K Brunnermeier and Stefan Nagel. Do wealth fluctuations generate time-varying risk aversion? micro-evidence on individuals asset allocation (digest summary). *American Economic Review*, 98(3):713–736, 2008.

O. H. Bustos and V. J. Yohai. Robust estimates for arma models. *Journal of the American Statistical Association*, 81:155–168, 1986.

E. Cantoni and E. Ronchetti. Resistant selection of the smoothing parameter for smoothing splines. *Statistics and Computing*, 11:141–146, 2001.

Ole F Christensen, Gareth O Roberts, and Martin Sköld. Robust markov chain monte carlo methods for spatial generalized linear mixed models. *Journal of Computational and Graphical Statistics*, 15(1):1–17, 2006.

T. Cipra. Robust exponential smoothing. *Journal of Forecasting*, 11:57–69, 1992.

T. Cipra and T Hanzak. Exponential smoothing for time series with outliers. *Kybernetika*, 4:165–178, 2011.

P. Cizek. Efficient robust estimation of time-series regression models. *Application of Mathematics*, 53:267–279, 2008.

Noel Cressie. Fitting variogram models by weighted least squares. *Journal of the International Association for Mathematical Geology*, 17(5):563–586, 1985.

C. Croux, S. Gelper, and K. Mahieu. Robust exponential smoothing of multivariate time series. *Journal of Computational Statistics & Data Analysis*, 54: 2999–3006, 2010.

X. de Luna and M. G. Genton. Robust simulation-based estimation of arma models. *Journal of Computational and Graphical Statistics*, 10(2):370–387, 2001.

L. Denby and R. D. Martin. Robust estimation of the first-order autoregressive parameter. *Journal of the American Statistical Association*, 74(365):140–146,

1979.

R. Fried, J. Einbeck, and U. Gather. Weighted repeated median smoothing and filtering. *Journal of the American Statistical Association*, 102:1300–1308, 2007.

S. Gelper, R. Fried, and C. Croux. Robust forecasting with exponential and Holt-Winters smoothing. *Journal of Forecasting*, 29:285–300, 2010.

M. G. Genton and E. Ronchetti. Robust indirect inference. *Journal of the American Statistical Association*, 98:1–10, 2003.

Marc G Genton. Highly robust variogram estimation. *Mathematical Geology*, 30(2):213–221, 1998.

C. W. Granger and M. J. Morris. Time Series Modelling and Interpretation. *Journal of the Royal Statistical Society. Series A*, pages 246–257, 1976.

S. Guerrier and R. Molinari. On the identifiability of latent models for dependent data. *Submitted manuscript*, 2016a.

S. Guerrier and R. Molinari. Wavelet variance for random fields: an m-estimation framework. *Submitted manuscript*, 2016b.

S. Guerrier, A. Waegli, J. Skaloud, and M. P. Victoria-Feser. Fault detection and isolation in multiple mems-imus configurations. *Aerospace and Electronic Systems, IEEE Transactions on*, 48(3):2015–2031, 2012.

S. Guerrier, R. Molinari, and M. P. Victoria-Feser. Estimation of Time Series Models via Robust Wavelet Variance. *Austrian Journal of Statistics*, pages 267–277, 2014.

Stéphane Guerrier, Jan Skaloud, Yannick Stebler, and Maria-Pia Victoria-Feser. Wavelet-variance-based estimation for composite stochastic processes. *Journal of the American Statistical Association*, 108(503):1021–1030, 2013.

Lars Peter Hansen. Large sample properties of generalized method of moments estimators. *Econometrica: Journal of the Econometric Society*, pages 1029–1054, 1982.

W. Härdle and T. Gasser. Robust non-parametric function fitting. *Journal of the Royal Statistical Society, Ser. B*, 46:42–51, 1984.

K. W. Hipel and A. I. McLeod. *Time series modelling of water resources and environmental systems*. Elsevier, 1994.

HK Iverson and RH Randles. The effects on convergence of substituting parameter estimates into u-statistics and other families of statistics. *Probability Theory and Related Fields*, 81(3):453–471, 1989.

H. Krim and I. Schick. Minimax description length for signal denoising and optimized representation. 45:898–908, 1999.

H Kunsch. Infinitesimal robustness for autoregressive processes. *The Annals of Statistics*, pages 843–863, 1984.

D. La Vecchia and F. Trojani. Infinitesimal robustness for diffusions. *Journal of the American Statistical Association*, 105:703–712, 2010.

Y. Ma and M. Genton. Highly robust estimation of the autocovariance function. *Journal of Time Series Analysis*, 21:663–684, 2000.

L. Mancini, E. Ronchetti, and F. Trojani. Optimal conditionally unbiased bounded-influence inference in dynamic location and scale models. *Journal of the American Statistical Association*, 100:628–641, 2005.

R. A. Maronna, R. D. Martin, and V. J. Yohai. *Robust Statistics: Theory and Methods*. Wiley, Chichester, West Sussex, UK, 2006.

R. D. Martin and V. J. Yohai. Influence functionals for time series. *The Annals of Statistics*, 14:781–818, 1986.

C. J. Masreliez and R. D. Martin. Robust Bayesian estimation for the linear model and robustifying the Kalman filter. *IEEE Transactions on Automatic Control*, 22:361–371, 1977.

F. F. Molinares, V. A. Reisen, and F. Cribari-Neto. Robust estimation in long-memory processes under additive outliers. *Journal of Statistical Planning and Inference*, 139:2511–2525, 2009.

Debashis Mondal and Donald B Percival. M-estimation of wavelet variance. *Annals of the Institute of Statistical Mathematics*, 64(1):27–53, 2012a.

Debashis Mondal and Donald B Percival. Wavelet variance analysis for random fields on a regular lattice. *Image Processing, IEEE Transactions on*, 21(2): 537–549, 2012b.

N. Muler, D. Pea, and V. J. Yohai. Robust estimation for ARMA models. *Annals of Statistics*, 37:816–840, 2009.

Whitney K Newey and Daniel McFadden. Large sample estimation and hypothesis testing. *Handbook of econometrics*, 4:2111–2245, 1994.

C. Ortelli and F. Trojani. Robust efficient method of moments. *Journal of Econometrics*, 128:6997, 2005.

Alan Pankratz. *Forecasting with dynamic regression models*, volume 935. John Wiley & Sons, 2012.

Donald B Percival. On estimation of the wavelet variance. *Biometrika*, 82(3): 619–631, 1995.

O. Renaud. Sensitivity and other properties of wavelet regression and density estimators. *Statistica Sinica*, 12:1275–1290, 2002.

E. Ronchetti and F. Trojani. Robust inference with GMM estimators. *Journal of Econometrics*, 101:37–69, 2001.

P. J. Rousseeuw and C. Croux. Alternatives to the median absolute deviation. *Journal of the American Statistical Association*, 88:1273–1283, 1993.

J. D. Salas and R. A. Smith. Physical basis of stochastic models of annual flows. *Water Resources Research*, 17(2):428–430, 1981.

S. Sardy, P. Tseng, and A. G. Bruce. Robust wavelet denoising. *IEEE Transactions on Signal Processing*, 49:1146–1152, 2001.

A. J. Q. Sarnaglia, V. A. Reisen, and C. Lévy-Leduc. Robust estimation of periodic autoregressive processes in the presence of additive outliers. *Journal of Multivariate Analysis*, 101:2168–2183, 2010.

Jiri Slacalek and Martin Sommer. What drives household saving? 2012.

Y. Stebler, S. Guerrier, J. Skaloud, and M. P. Victoria-Feser. The generalised method of wavelet moments for inertial navigation filter design. *IEEE Transactions on Aerospace and Electronic Systems*, 2014. To appear.

Kevin Thon, Marc Geilhufe, and Donald B Percival. A multiscale wavelet-based test for isotropy of random fields on a regular lattice. *Image Processing, IEEE Transactions on*, 24(2):694–708, 2015.

David Titterton and John L Weston. *Strapdown inertial navigation technology*,

volume 17. IET, 2004.

A Sezin Tokar and Peggy A Johnson. Rainfall-runoff modeling using artificial neural networks. *Journal of Hydrologic Engineering*, 4(3):232–239, 1999.

J. W. Tukey. *Exploratory Data Analysis*. Addison-Wesley, Massachussets, 1977. Preliminary editions: 1970, 1971.

Julio Videras and Stephen Wu. The behavioral effects of financial shocks. 2004.

V. J. Yohai and R. H. Zamar. High breakdown point estimates of regression by means of the minimization of an efficient scale. *Journal of the American Statistical Association*, 83:406–413, 1988.

**Appendix A: Short literature review**

A detailed discussion on robust estimation and inference methods for time series models can be found in Maronna et al. (2006), Chapter 8. Most of the literature in this domain has dealt with standard time series models such as autoregressive and/or moving average models. Kunsch (1984) proposes optimal robust estimators of the parameters of autoregressive processes by studying the properties of their influence function (see also Martin and Yohai, 1986). Denby and Martin (1979) develop a generalized $M$-estimator for the parameter of a first-order autoregressive process whereas Bustos and Yohai (1986), Allende and Heiler (1992) and de Luna and Genton (2001) extend the research to include moving average models using generalized $M$-estimation theory and indirect inference (see e.g. **?**). Bianco et al. (1996) propose a class of robust estimators for regression models with ARIMA errors based on $\tau$-estimators of scale (Yohai and Zamar, 1988). Ronchetti and Trojani (2001) develop a robust version of the generalized method of moments (proposed by Hansen, 1982) for estimating the parameters of time series models in economics, and Ortelli and Trojani (2005) further develop a robust efficient method of moments. Mancini et al. (2005) propose optimal bias-robust estimators for a class of conditional location and scale time series models while La Vecchia and Trojani (2010) develop conditionally unbiased optimal robust estimators for general diffusion processes, for which approximation methods for computing integrals are needed. Cizek (2008) studies the properties of a two-step least weighted squares robust time-series regression estimator and Agostinelli and Bisaglia (2010) propose a weighted maximum likelihood estimator for ARFIMA processes, for which Molinares et al. (2009) propose an alternative estimator under additive outliers. Sarnaglia et al. (2010) suggest a robust estimation procedure for the parameters of the periodic AR model as an extension of the robust scale and covariance functions given in, respectively, Rousseeuw and Croux (1993) and Ma and Genton (2000).

Another means to obtain robust estimators for the parameters of a time series model when it can be written as a state-space model is by means of robust (Kalman) filtering. Robustification of the Kalman filter was originated with Masreliez and Martin (1977) and Cipra (1992) who propose robust modifications of exponential smoothing (see also Cipra and Hanzak, 2011 and Croux et al., 2010 for a multivariate version). For a robust version of the Holt-Winters smoother, see Gelper et al. (2010). Muler et al. (2009) develop a class of robust estimates for ARMA models that are closely related to robust filtering. Robust filtering can also possibly provide a way to robustly estimate the WV (although stronger assumptions on the unerlying model would have to be made). However, in this case, the only attempt to studying the robustness properties of wavelet filtering has been made in the identically and independently distributed (iid) case where Renaud (2002) develops, among others, the IF of the Haar-based wavelet coefficients and concludes that the IF depends on the location of the contaminated data with respect to the dyadic grid and can be infinite. As in the case of the wavelet coefficients, many classical filtering methods are unbounded and for this reason several robust local filters have been proposed so far since the median filter proposal from Tukey (1977): Bruce et al. (1994) pre-process the estimation of the wavelet coefficients via a "fast and robust smooth/cleaner"; Krim and Schick (1999) derive a robust estimator of the wavelet coefficients based on minimax description length; Härdle and Gasser (1984) develop a locally weighted smoothing using $M$-estimation and Fried et al. (2007) propose a non-parametric, weighted repeated median filter. Sardy et al. (2001) propose a robust wavelet-based estimator using a robust loss-penalized function, for which appropriately choosing the smoothing parameter is an important robustness issue as revealed, for example, by Cantoni and Ronchetti (2001).

The literature specifically on robust estimation of spatial models is less abundant. An overview of some robust methods which are not necessarily directly linked to robust spatial model estimation can be found in Anselin (2013). In Christensen et al. (2006) a robust Markov-Chain approach is discussed for spatial generalized linear mixed models whereas a few methods have been proposed to robustly estimate the variogram (or semi-variogram) of a spatial process and some examples can be found in Cressie (1985) or Genton (1998). Ideally, these can be used as robust auxiliary parameters in a minimum-distance estimator approach. Literature dealing with statistical robustness in the estimation of higher-dimensional models is even more scarce, if not nonexistent, to the best of our knowledge.

**Appendix B: RGMWM Asymptotic Properties**

The results of the proofs in this appendix largely follow the conditions and steps for standard extremum estimators (see Newey and McFadden, 1994). However, since we consider cases where we let $J \rightarrow \infty$ (i.e. the number of moment conditions goes to infinity with the sample size), we nevertheless illustrate the steps and details which allow these standard conditions to hold also in the RGMWM setting. In the following proofs $\|\cdot\|$ indicates the Frobenius norm if the object is a matrix.

*B.1. RGMWM Consistency*

*Proof.* Given conditions **(C1)** to **(C3)**, let $Q(\boldsymbol{\theta}) \equiv (\boldsymbol{\nu}(\boldsymbol{\theta}_0) - \boldsymbol{\nu}(\boldsymbol{\theta}))^T \boldsymbol{\Omega}(\boldsymbol{\nu}(\boldsymbol{\theta}_0) - \boldsymbol{\nu}(\boldsymbol{\theta})) = \|\boldsymbol{\nu}(\boldsymbol{\theta}_0) - \boldsymbol{\nu}(\boldsymbol{\theta})\|_{\boldsymbol{\Omega}}^2$ and $Q_N(\boldsymbol{\theta}) \equiv (\hat{\boldsymbol{\nu}} - \boldsymbol{\nu}(\boldsymbol{\theta}))^T \hat{\boldsymbol{\Omega}} (\hat{\boldsymbol{\nu}} - \boldsymbol{\nu}(\boldsymbol{\theta})) = \|\hat{\boldsymbol{\nu}} - \boldsymbol{\nu}(\boldsymbol{\theta})\|_{\hat{\boldsymbol{\Omega}}}^2$ and define $\boldsymbol{\Omega}^* = \hat{\boldsymbol{\Omega}} - \boldsymbol{\Omega}$. By Theorem 2.1. of Newey and McFadden (1994) we want to prove that $Q_N(\boldsymbol{\theta})$ converges uniformly in probability to $Q(\boldsymbol{\theta})$. By the triangular inequality and Cauchy-Schwartz we have that

$$|Q_N(\boldsymbol{\theta}) - Q(\boldsymbol{\theta})| \leq \left| \underbrace{\|\boldsymbol{\nu}(\boldsymbol{\theta}_0) - \boldsymbol{\nu}(\boldsymbol{\theta})\|_{\boldsymbol{\Omega}^*}^2}_{a_1} \right| + \left| \underbrace{\|\hat{\boldsymbol{\nu}} - \boldsymbol{\nu}(\boldsymbol{\theta}_0)\|_{\hat{\boldsymbol{\Omega}}}^2}_{a_2} \right| + \left| \underbrace{2 (\boldsymbol{\nu}(\boldsymbol{\theta}_0) - \boldsymbol{\nu}(\boldsymbol{\theta})^T \hat{\boldsymbol{\Omega}} (\hat{\boldsymbol{\nu}} - \boldsymbol{\nu}(\boldsymbol{\theta}_0)))}_{a_3} \right|.$$

Considering term $a_1$ and using the same inequalities, we have

$$a_1 \leq \|\boldsymbol{\nu}(\boldsymbol{\theta}_0) - \boldsymbol{\nu}(\boldsymbol{\theta})\|^2 \|\boldsymbol{\Omega}^*\| = \sum_{j=1}^{J} (\nu_j(\boldsymbol{\theta}_0) - \nu_j(\boldsymbol{\theta}))^2 \sqrt{\sum_{i=1}^{J} \sum_{j=1}^{J} (\hat{\omega}_{i,j} - \omega_{i,j})^2}.$$

By **(C1)** and **(C6)** we have that $(\nu_j(\boldsymbol{\theta}_0) - \nu_j(\boldsymbol{\theta}))^2$ is bounded by a quantity $B$ and

$$\sqrt{\sum_{i=1}^{J} \sum_{j=1}^{J} (\tilde{\omega}_{i,j} - \omega_{i,j})^2} \leq J \sqrt{(\tilde{\omega}_{i,j} - \omega_{i,j})^2} = \mathcal{O}_p \left( \frac{J}{g(N)} \right)$$

thereby giving

$$a_1 \leq JB \, \mathcal{O}_p \left( \frac{J}{g(N)} \right) = \mathcal{O}_p \left( \frac{J^2}{g(N)} \right)$$

which, by **(C6)**, tends to 0 in probability. Moreover, using the results in Guerrier and Molinari (2016b), we have that

$$a_2 \leq \|\hat{\boldsymbol{\Omega}}\| \|\hat{\boldsymbol{\nu}} - \boldsymbol{\nu}(\boldsymbol{\theta}_0)\|^2 = \mathcal{O}_p \left( \frac{J}{g(N)} \right) \mathcal{O}_p \left( \frac{J}{\sqrt{N_J}} \right) = \mathcal{O}_p \left( \frac{J}{\min(g(N), \sqrt{N_J})} \right).$$

Finally, we have

$$a_3 \leq 2\,\|\hat{\boldsymbol{\Omega}}\|\,\|\boldsymbol{\nu}(\boldsymbol{\theta}_0) - \boldsymbol{\nu}(\boldsymbol{\theta})\|\,\|\hat{\boldsymbol{\nu}} - \boldsymbol{\nu}(\boldsymbol{\theta}_0)\|.$$

By **(C6)** we have that $\|\hat{\boldsymbol{\Omega}}\| \leq \lambda$ and by **(C1)** we have $\|\boldsymbol{\nu}(\boldsymbol{\theta}_0) - \boldsymbol{\nu}(\boldsymbol{\theta})\| = \mathcal{O}_p(\sqrt{J})$. Moreover, given the results in Guerrier and Molinari (2016b) we have $\|\hat{\boldsymbol{\nu}} - \boldsymbol{\nu}(\boldsymbol{\theta}_0)\| = \mathcal{O}_p(\sqrt{J/N_J})$ which gives

$$a_3 \leq \lambda\,\mathcal{O}_p(\sqrt{J})\,\mathcal{O}_p\left(\sqrt{\frac{J}{N_J}}\right) = \mathcal{O}_p\left(\frac{J}{\sqrt{N_J}}\right).$$

Therefore, we have that

$$\sup_{\boldsymbol{\theta}\in\boldsymbol{\Theta}}\bigl|Q_N(\boldsymbol{\theta}) - Q(\boldsymbol{\theta})\bigr| \xrightarrow{\mathcal{P}} 0.$$

Based on this and using conditions **(C1)**, **(C3)**, **(C4)**, we finally have

$$\hat{\boldsymbol{\theta}} \xrightarrow{\mathcal{P}} \boldsymbol{\theta}_0$$

thus concluding the proof. $\square$

### B.2. RGMWM Asymptotic Normality

*Proof.* Given the results on the consistency in Proposition 2.1, the proof of asymptotic normality of $\hat{\boldsymbol{\theta}}$ naturally follows the standard proof of asymptotic normality for GMM estimators. Here we therefore simply give a quick overview of the steps which are based on existing results (see Newey and McFadden, 1994). Let $Q(\boldsymbol{\theta}) \equiv (\boldsymbol{\nu}(\boldsymbol{\theta}_0) - \boldsymbol{\nu}(\boldsymbol{\theta}))^T \boldsymbol{\Omega}(\boldsymbol{\nu}(\boldsymbol{\theta}_0) - \boldsymbol{\nu}(\boldsymbol{\theta})) = \|\boldsymbol{\nu}(\boldsymbol{\theta}_0) - \boldsymbol{\nu}(\boldsymbol{\theta})\|^2_{\boldsymbol{\Omega}}$ and $Q_N(\boldsymbol{\theta}) \equiv (\hat{\boldsymbol{\nu}} - \boldsymbol{\nu}(\boldsymbol{\theta}))^T \hat{\boldsymbol{\Omega}}(\hat{\boldsymbol{\nu}} - \boldsymbol{\nu}(\boldsymbol{\theta})) = \|\hat{\boldsymbol{\nu}} - \boldsymbol{\nu}(\boldsymbol{\theta})\|^2_{\hat{\boldsymbol{\Omega}}}$. Moreover, let us define $\boldsymbol{\Omega}_N = 1/2(\hat{\boldsymbol{\Omega}} + \hat{\boldsymbol{\Omega}}^T)$. Finally, using condition **(C8)**, let us denote $B_N(\boldsymbol{\theta}) = \partial/\partial\boldsymbol{\theta}\, Q_N(\boldsymbol{\theta})$, with $B(\boldsymbol{\theta})$ being the counterpart for $Q(\boldsymbol{\theta})$, $H_N(\boldsymbol{\theta}) = \partial/\partial\boldsymbol{\theta}^T B_N(\boldsymbol{\theta})$ and $D(\boldsymbol{\theta}) = \partial/\partial\boldsymbol{\theta}\, \boldsymbol{\nu}(\boldsymbol{\theta})$. Since we have that $Q_N(\hat{\boldsymbol{\theta}})$ minimizes $Q_N(\boldsymbol{\theta})$ by definition, we consequently have that

$$B_N(\hat{\boldsymbol{\theta}}) = 0.$$

By taking a Maclaurin expansion of $B_N(\hat{\boldsymbol{\theta}})$ around $\boldsymbol{\theta}_0$ we have

$$B_N(\hat{\boldsymbol{\theta}}) = B_N(\boldsymbol{\theta}_0) + H_N(\boldsymbol{\theta}^*)(\hat{\boldsymbol{\theta}} - \boldsymbol{\theta}_0) \tag{A-1}$$

where $\|\boldsymbol{\theta}^* - \boldsymbol{\theta}_0\|^2 \leq \|\hat{\boldsymbol{\theta}} - \boldsymbol{\theta}_0\|^2$. Using the two equalities, rearranging and multiplying by $\sqrt{N_J}$ give us

$$\sqrt{N_J}(\hat{\boldsymbol{\theta}} - \boldsymbol{\theta}_0) = (-H_N(\boldsymbol{\theta}^*)^{-1})(\sqrt{N_J}B_N(\boldsymbol{\theta}_0)).$$

Let us focus on term $H_N(\boldsymbol{\theta}^*)$ and let us now take a Taylor expansion of $B_N(\hat{\boldsymbol{\theta}})$ around $\boldsymbol{\theta}_0$ which gives

$$B_N(\hat{\boldsymbol{\theta}}) = B_N(\boldsymbol{\theta}_0) + H_N(\boldsymbol{\theta}_0)(\hat{\boldsymbol{\theta}} - \boldsymbol{\theta}_0) + \mathcal{O}(\|\hat{\boldsymbol{\theta}} - \boldsymbol{\theta}_0\|^2).$$

Taking the difference with the Maclaurin expansion in (A-1) implies that

$$\|H_N(\boldsymbol{\theta}^*) - H(\boldsymbol{\theta}_0)\| = \mathcal{O}_p(\|\hat{\boldsymbol{\theta}} - \boldsymbol{\theta}_0\|)$$

and hence, by continuity of matrix inversion, we have

$$-H_N(\boldsymbol{\theta}^*)^{-1} \xrightarrow{\mathcal{P}} H(\boldsymbol{\theta}_0)^{-1}.$$

Using **(C7)** to **(C8)**, we finally obtain

$$\sqrt{N_J}\left(\hat{\boldsymbol{\theta}} - \boldsymbol{\theta}_0\right) \xrightarrow[N_J \to \infty]{\mathcal{D}} \mathcal{N}\left(\mathbf{0}, \mathbf{BVB}^T\right)$$

with $\mathbf{B} = \mathbf{H}(\boldsymbol{\theta}_0)^{-1}\mathbf{D}(\boldsymbol{\theta}_0)^T\boldsymbol{\Omega}$, thus concluding the proof. □

## Appendix C: Additional Results from Simulation Studies

Here we report some additional results from the simulation studies made in Section 3. Table 6 gives the estimation times in seconds and convergence rates of the RGMWM and INDI estimators under an uncontaminated setting.

|  | Model | Sample size | Median | Conv. rate (%) |
|---|---|---|---|---|
| RGMWM | **ARMA(1,2)** | 1,000 | $5.3 \cdot 10^{-1}$ | 100 |
|  | **ARMA(3,1)** | 1,000 | $7.67 \cdot 10^{-1}$ | 100 |
|  | **SSM** | 100 | $5.6 \cdot 10^{-2}$ | 100 |
| INDI | **ARMA(1,2)** | 1,000 | $1.473 \cdot 10^{2}$ | 100 |
|  | **ARMA(3,1)** | 1,000 | $1.361 \cdot 10^{2}$ | 99.8 |
|  | **SSM** | 100 | $1.025 \cdot 10^{2}$ | 29.2 |

TABLE 6
*Sample size, median computational time in seconds and convergence rates of the RGMWM and INDI estimators for the models **ARMA(1,2)**, **ARMA(3,1)** and **SSM** in an uncontaminated setting.*

## Appendix D: Additional Results from Application on Inertial Sensor Data

In this appendix we find other results for the application on the inertial sensor measurements studied in Section 4.1. Figure 6 shows the WV plots which are in the same spirit of those provided for the application in Section 4.2 and are interpreted in the same manner.

It can be seen how the different estimations are significantly different at the lower scales where the outliers have more of an influence on estimations. The model-implied WV is plotted based on the parameter estimates of the GMWM and RGMWM which can be found in Table 7 along with their respective confidence intervals. Given the length of the signal, it can be seen how some estimates are the same between methods and the confidence intervals are extremely tight, appearing to be identical to the estimate itself due to the rounding (see, for example, the estimates and confidence intervals for $\rho_1$ and $\rho_2$). Considering these details, we can remark that the parameters of the first two autoregressive processes are significantly different from each other underlining that although the percentage of outliers is considerably low, the contamination appears to have an impact on the estimation process and robust methods should therefore be preferred.

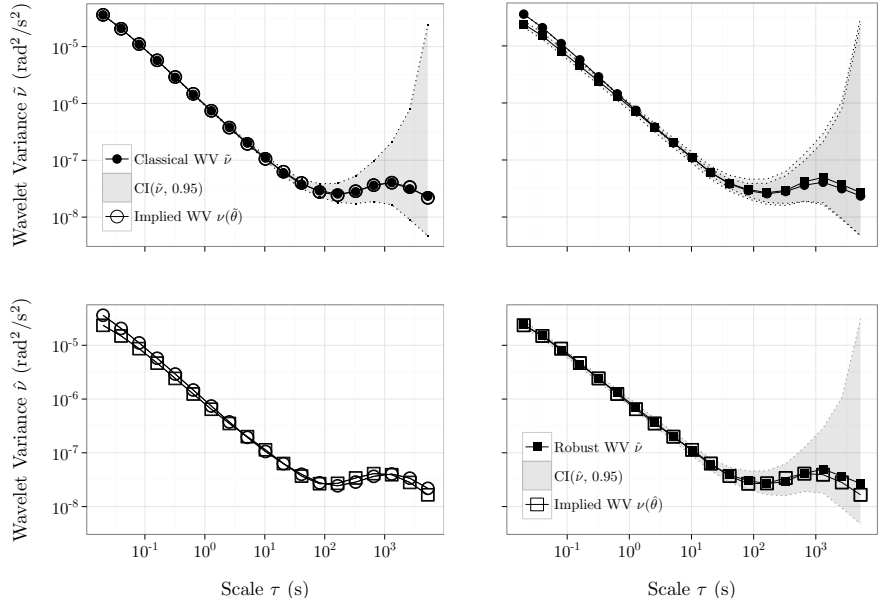

FIG 6. *Log-log scale Wavelet Variance plots for the inertial sensor series. Top left: classic estimated WV superposed with model-implied WV based on the parameters estimated through the GMWM. Top right: classic and robust estimated WV with respective confidence intervals superposed. Bottom left: classic and robust model-implied WV based on GMWM and RGMWM estimates respectively. Bottom right: robust estimated WV superposed with model-implied WV based on the parameters estimated through the RGMWM.*

TABLE 7
*State-space model estimates for the gyroscope data in static conditions. Estimated parameters with GMWM and RGMWM estimators with $\rho_i$ being the $i^{th}$ autoregressive parameter, $v_i^2$ the innovation variance of the $i^{th}$ autoregressive model. Confidence intervals (CI) based on the approach used in Guerrier et al. (2013).*

| | GMWM | | RGMWM | |
| --- | --- | --- | --- | --- |
| | Estimate | CI($\cdot$, 95%) | Estimate | CI($\cdot$, 95%) |
| $\rho_1$ | $8.9546 \cdot 10^{-2}$ | $(8.9546 \cdot 10^{-2}\ ;\ 8.9546 \cdot 10^{-2})$ | $1.4816 \cdot 10^{-1}$ | $(1.4816 \cdot 10^{-1}\ ;\ 1.4816 \cdot 10^{-1})$ |
| $v_1^2$ | $7.8760 \cdot 10^{-5}$ | $(7.8612 \cdot 10^{-5}\ ;\ 7.8909 \cdot 10^{-5})$ | $5.5325 \cdot 10^{-5}$ | $(5.5211 \cdot 10^{-5}\ ;\ 5.5439 \cdot 10^{-5})$ |
| $\rho_2$ | $9.9831 \cdot 10^{-1}$ | $(9.9831 \cdot 10^{-1}\ ;\ 9.9831 \cdot 10^{-1})$ | $9.9687 \cdot 10^{-1}$ | $(9.9687 \cdot 10^{-1}\ ;\ 9.9687 \cdot 10^{-1})$ |
| $v_2^2$ | $2.5632 \cdot 10^{-10}$ | $(2.0887 \cdot 10^{-10}\ ;\ 3.0377 \cdot 10^{-10})$ | $1.0466 \cdot 10^{-9}$ | $(9.4291 \cdot 10^{-10}\ ;\ 1.1504 \cdot 10^{-9})$ |
| $\rho_3$ | $9.9997 \cdot 10^{-1}$ | $(9.9995 \cdot 10^{-1}\ ;\ 9.9999 \cdot 10^{-1})$ | $9.9997 \cdot 10^{-1}$ | $(9.9995 \cdot 10^{-1}\ ;\ 9.9999 \cdot 10^{-1})$ |
| $v_3^2$ | $1.1915 \cdot 10^{-11}$ | $(7.5299 \cdot 10^{-12}\ ;\ 1.6300 \cdot 10^{-11})$ | $1.3626 \cdot 10^{-11}$ | $(8.7412 \cdot 10^{-12}\ ;\ 1.8511 \cdot 10^{-11})$ |